# An expanded version of the front door criterion under the potential outcome framework: A guide to empiricist

Zexuan Chen[a]*

[a] *Department of Archaeology, Durham University, Durham, United Kingdom*

zexuan.chen@durham.ac.uk

# An expanded version of the front door criterion under the potential outcome framework: A guide to empiricist

## ABSTRACT

In recent years, the front-door criterion (FDC) has been increasingly noticed in economics. However, using regression to apply the FDC under the potential outcome framework (RCM) is not purely the same as the original non-parametric method in the structure causal model (SCM). This article aims to exposit the vital additional factors of using the FDC under the RCM. It first defines a new type of causal effect: path average causal effect (PATE), which is used for further exposition. Then, the key assumptions of the FDC are redefined with the language of the RCM. Four further assumptions are made specifically for the FDC under the RCM framework. After that, the causal connotations of the FDC estimates are elaborated in detail with PATE, and the estimation bias caused by violating some new assumptions is theoretically derived. Rigorous simulation data are used to confirm the theoretical derivation. It is proved that the FDC can still provide useful insights into causal relationships even when some key assumptions are violated. Finally, the FDC is also comprehensively compared with the instrumental variables (IV). The analyses of this paper prove that the FDC can provide new insights into causal relationships compared with the conventional methods in economics.

## 1. Introduction

The front-door criterion (FDC), proposed by Pearl in 1995 (Pearl (1995, 2009)), is a powerful tool for causal effect estimation. Unlike the backdoor criterion (BDC), which strongly relies on the causal diagram, the FDC does not require domain knowledge of the overall causal diagram. This makes it easier to use and more acceptable to empirical researchers as whether the true causal relationships can be represented by a directed acyclic graph (DAG) has always been a focus of controversy.

Although the FDC has been discussed by economists for more than 15 years (e.g., Cohen and Malloy (2014), Imbens (2020), Gupta, Lipton, and Childers (2020)), it has only come into use until 2024. In current social science and economics, only very limited studies adopt the FDC (e.g., Glynn and Kashin (2017, 2018), Bellemare, Bloem, and Wexler (2024)). The method to apply the FDC by economists is a two-step regression (Bellemare, Bloem, and Wexler (2024)).

However, some problems arise when economists apply the FDC by the two-step regression, which does not exist when it serves as a nonparametric method in the structure causal model (SCM). This paper aims to exposit these further factors that must be considered for applying the FDC in the potential outcome model (RCM) and compare it with the instrumental variables (IV). In section 2, I briefly introduce the FDC under the framework of the SCM and the RCM (i.e., the two-step regression method). In section 3, I define a new type of causal effect: path average causal effect, and redefine the assumptions of the FDC with the language of the RCM. The total of seven assumptions are more comprehensive than the previous assumptions in the SCM and are vital to guarantee the reliability of the FDC when applying it by the two-step regression. I then explain the causal connotations of FDC estimates in detail with the PATE. More importantly, the theoretical derivation shows that FDC can still provide useful insights into causal relationships even when some key assumptions are violated, which is also confirmed by the simulated data tests in section 5. Therefore, the empirical researcher does not need to strictly adhere to all assumptions when using it. I also make comprehensive comparisons between it and the IV in section 4. These analyses prove the value of the FDC and help empirical researchers gain a more comprehensive understanding of it. The FDC has the potential to become a more commonly used tool in the empirical toolkit.

## 2. The front-door criterion in the structure causal model

The insight of FDC is that if there exists a set of mediator variables M, which satisfy three conditions relative to an ordered pair of variables (X, Y), we can then use M for adjustment of confounders if P(x, m)>0 and three conditions are met.

Condition 1: M intercepts all directed paths from X to Y.

Condition 2: There is no backdoor path from X to M.

Condition 3: All backdoor paths from M to Y are blocked by X.

The first condition implies that X can only affect Y through M. Conditions 2 and 3 suggests that there are no confounding factors between X and M, as well as Y and M. If all three assumptions are met, the front-door adjustment formula is given by

$$P\big(y\big|do(x)\big) = \sum_{m} P(m|x) \sum_{x'} P(y|x', m)P(x'). \quad (1)$$

The FDC is a nonparametric method. However, it can also be linked to the regression models serving as a parametric method. Assuming linear relationships exist between X, M, and Y, the FDC is conducted by a two-step regression (Bellemare, Bloem, and Wexler (2024)).

$$M_i = \alpha + \beta X_i + \varepsilon_i. \quad (2)$$

$$Y_i = \gamma + \delta M_i + \varphi X_i + \omega_i. \quad (3)$$

Equation (2) estimates the causal effect between treatment variables and mediator variables if Condition 2 is satisfied. Equation (3) conducts the causal estimates between outcome variables and mediator variables under the premise of Condition 3. Then, given Condition 1 and linear relationships between all variables, the causal effect of X on Y is calculated by multiplying the coefficient estimates $\hat{\beta}$ and $\hat{\delta}$.

## 3. The front-door criterion in the potential outcome model

In this section, I first define some causal effects and the critical assumptions for the front-door criterion in a more detailed and comprehensive way using the language of the potential outcome framework. Then, the causal connotation of the FDC estimates is interpreted. Finally, explain how biased the estimation will be if some key assumptions are violated.

### *3.1 Key Definitions and Assumptions*

Here, I first give the definitions of some causal effects that will be discussed further in this article. Then, define the key assumptions for this method.

Consider a binary treatment and mediator, then the definition of causal effect can be given.

Definition 1: Causal Effect on Individual Unit (ITE).

Say $Y_i$ and $X_i$ as outcome and treatment variables for individual unit i. The causal effect of X on Y for that individual is $Y_i(1) - Y_i(0)$.

Definition 2: Average Causal Effect on Population (ATE).

Say $Y_i$ and $X_i$ as outcome and treatment variables for individual unit i. The average causal effect of X on Y for the population is $E[Y_i(1) - Y_i(0)]$.

Definition 3: Local Average Causal Effect on Local individuals (LATE) (Imbens and Angrist (1994), Angrist, Imbens, and Rubin (1996), Angrist and Pischke (2009)).

Say $Y_i$ and $X_i$ as outcome and treatment variables for individual unit i. There exists a local population, whose treatment variables are driven by other factors (e.g., instrumental variables). The average causal effect of X on Y for this local population is $E[Y_i(1) - Y_i(0) | X_i(1) - X_i(0) = 1]$. The causal effect is only estimated for local individuals whose $X_i(1) - X_i(0) = 1$.

Definition 4: Path Causal Effect on Individual Unit (PITE)

This paper defines a new type of causal effect: PITE. That is a causal effect of an individual unit from a single path. Take the path X→M→Y as an example. Say $Y_i$ and $X_i$ as outcome and treatment variables for individual unit i. The path through a mediator variable M is applied for this individual, and M is the only path from X to Y. The causal effect of X on Y for that individual through the path is $\left(Y_i(1) - Y_i(0)\right) \cdot \left(M_i(1) - M_i(0)\right)$ (see also equation (4)). The causal effect is estimated for the individual i through the path X→M→Y. PITE=ITE, if X→M→Y is the only path.

Definition 5: Path Average Causal Effect on Population (PATE)

The causal effect of the whole population from a single path. Take the path X→M→Y as an example. Say $Y_p$ and $X_p$ as outcome and treatment variables for individual units whose treatment variables positively affect their mediator variables $M_p$. While $Y_n$ and $X_n$ are outcome and treatment variables for individual units whose treatment variables negatively affect their mediator variables $M_n$. $P(p)$ is the proportion of individuals of p compared to the whole population and $P(n)$ is the proportion of individuals of n. The path through a mediator variable M is applied for the whole population, and M is the only path from X to Y. The average causal effect of X on Y for the population through the path is $E\left[Y_p(1) - Y_p(0)\right] \cdot P(p) - E[Y_n(1) - Y_n(0)] \cdot P(n)$. (see also equation (5) and (6)). The causal effect is estimated for the whole population through the path X→M→Y. PATE=ATE, if X→M→Y is the only path.

Assumption1: Identifiability.

The average causal effect between X and M, Y and M is not zero. Taking the binary treatment and mediator as an example $(\forall x, m \in \{0,1\})$, that is $E[M_i(1) - M_i(0)] \neq 0$, $E[Y_i(1) - Y_i(0)] \neq 0$.

This assumption makes the causal effect identifiable. It is a stronger version of the original assumption (P(x, m)>0). If the causal effect is zero, then no results can be given. This means the mediator variable chosen is an invalid one. In such a case, we cannot use this method because the mediator variable chosen does not represent the true mediation in the real world. There is also another possibility. That is there is indeed no causal relationship between X and Y. In such a case, concluding that there is no causal relation between our interesting variables is fine. However, we can hardly distinguish between these two cases. This part is quite a black box, which is also the restriction of this method (but all causal inference methods contain similar issues: a black box and story-telling style). It heavily relies on our domain knowledge, especially when we lack data, not suitable for conducting causal discovery. If this is the case (the causal effect is zero), other methods should be considered. Using multiple methods to compare the results will provide more solid conclusions.

Assumption 2: Unconfoundedness.

$$X \perp\!\!\!\perp M$$

$$Y \perp\!\!\!\perp M|X$$

This assumption means that there is no hidden variable that confounds treatment variables and mediator variables, and mediator variables and outcome variables are not confounded if we condition on treatment variables. If we use the linear model to illustrate this, then $E[X_i \cdot \varepsilon_i] = 0$ from equation (2) guarantees the coefficient of first-step regression catches the causal effect between X and M. While, $E[M_i \cdot \omega_i] = 0$ from equation (3) proves the coefficient of second-step regression reflects the causal effect between Y and M. If such a confounding variable does exist, it can be adjusted if it is observed. The biased estimation will be made if this assumption is not satisfied and the confounding variables cannot be observed.

Assumption 3: Stable Unit Treatment Value Assumption (SUTVA) (Rubin (1978, 1980, 1990)).

In our case, if $X_i = X_i'$, $M_i(X_i) = M_i(X_i')$.

If $X_i = X_i'$ and $M_i = M_i'$, $Y_i(M_i, X_i) = Y_i(M_i', X_i')$

SUTVA is a basic rule in the potential outcome framework, implying that potential outcomes for each unit are unaffected by the treatment status of others. This is an important restriction. All causal inference methods must obey this rule. For example, if we use Difference-in-differences (DID) to study policy impact. If the individuals from control groups migrate to the city in the treatment group after policy because of its positive effect. Then, SUTVA is violated. The migration adds additional heterogeneity for two groups and thus invalidates the method.

Assumption 4: Uniqueness.

$Y(X, M) = Y(X', M)$ for all $X$, $X'$ for all $M$.

This means for all individual units, no matter what values the treatment variable takes, the values of the outcome variable will be the same if the mediator variables take the same values. That is the mediator variables catch all the effects from X to Y (i.e., M is the unique path from X to Y). If this is not the case, we may fail to estimate the ATE. The reason is that if X can affect Y through other paths, then even if all confounding factors are removed, the relationship between Y and M will not be clear.

Assumption 5: Universality. That is all individual units i follow the path X→M→Y.

If only a part of the units follows the path through M, our estimates will be LATE (similar to that of the IV). In this case, if we cannot distinguish which individuals follow the path based on M, we may provide biased estimates (Heckman (1990), Angrist and Imbens (1991), Manski (1994)).

Assumption 6: No Heterogeneity.

If the function between X and M is different for some subgroups, the function between M and Y for all subgroups must be consistent. Take a binary case as an example. The population i consists of two groups. $M_p(1) - \mathrm{M}_p(0) = 1, M_n(1) - \mathrm{M}_n(0) = -1$ for subgroups p and n, respectively. Then, $E[Y_i(1) - Y_i(0)] = E\left[Y_p(1) - Y_p(0)\right] = E[Y_n(1) - Y_n(0)]$.

Monotonicity (Imbens and Angrist (1994), Angrist, Imbens, and Rubin (1996)) is not required for the FDC. No matter the positive or negative effect of treatment variables on mediator variables, it is just the truth and can be conveyed to the outcome variable. This is determined by the nature of the mediator variables. However, no heterogeneity must be followed, or the biased estimation will be made (see discussion in 3.2).

Assumption 7: Confounding Factors Distribution Assumption.

All confounding factors must follow the Normal Distribution $X \sim N(0, \sigma^2)$.

In linear systems, we assume that the error terms are all subject to a normal distribution with an expected value of 0. For the front-door criterion, the FDC variables may be affected by an exogenous variable U. These factors will make the FDC variable a collider in the causal diagram. If we condition on it, X and U are conditionally dependent (Pearl (2009), Pearl, Glymour, and Jewell (2016), Pearl and Mackenzie (2018)). It will introduce new confounding factors, making it difficult to recognize the true causal effects. Specifically, M is the result of the joint effect by X and U. Therefore, the second-stage regression obtains the effect of M on Y driven by both X and U, rather than the effect of M on Y driven by X alone. However, if U follows the assumption of a normal distribution with an expected value of 0, then these two causal effects will be equivalent in linear systems. If we isolate a part of M, which is only related to X, and regress it with Y, the regression coefficient will be the same as regressing Y and actual M. The only difference is in error terms.

Assumption 1 is a stronger version of the original FDC. Assumptions 2 and 4 rewrite the assumption in the FDC in the language of the potential outcome framework. Assumptions 3, 5, 6, and 7 are newly added to make the method more comprehensive and rigorous

### *3.2 Interpreting the Causal Connotations*

Consider a binary treatment and mediator $(\forall x, m \in \{0,1\})$, if all six assumptions above are met, the PITE of X on Y of the FDC for the population i is estimated by

$$
\begin{aligned}
&PITE \\
&= Y_i(1, M_i(1)) - Y_i(0, M_i(0)) \\
&= Y_i(M_i(1)) - Y_i(M_i(0)) \\
&= [Y_i(1) \cdot M_i(1) + Y_i(0) \cdot (1 - M_i(1))] - [Y_i(1) \cdot M_i(0) + Y_i(0) \cdot (1 - M_i(0))] \\
&\qquad = \underbrace{(Y_i(1) - Y_i(0))}_{\text{Path: M} \rightarrow \text{Y}} \cdot \underbrace{(M_i(1) - M_i(0))}_{\text{Path: X} \rightarrow \text{M}}. \qquad (4)
\end{aligned}
$$

Thus, in our method, the causal effect of an individual is composed of the causal effect of M on Y as well as the effect of X on M. $Y_i(1) - Y_i(0)$ follows the path M → Y. $M_i(1) - M_i(0)$ follows the path X → M.

Then, we can calculate the PATE.

$$
\begin{aligned}
&PATE \\
&= E[(Y_i(1, M_i(1)) - Y_i(0, M_i(0)))] \\
&= E[(Y_i(1) - Y_i(0)) \cdot (M_i(1) - M_i(0))] \\
&= E[(Y_i(1) - Y_i(0)|M_i(1) - M_i(0) = 1] \cdot P(M_i(1) - M_i(0) = 1) - E[(Y_i(1) \\
&\qquad - Y_i(0)|M_i(1) - M_i(0) = -1] \cdot P(M_i(1) - M_i(0) = -1). \qquad (5)
\end{aligned}
$$

Say $M_p(1) - M_p(0) = 1$ for positive individual units p, and $M_n(1) - M_n(0) = -1$ for negative individual units. $P(p)$ is the proportion of individuals of p compared to the whole population i and $P(n)$ is the proportion of individuals of n.

$$PATE = E[Y_p(1) - Y_p(0)] \cdot P(p) - E[Y_n(1) - Y_n(0)] \cdot P(n). \qquad (6)$$

Consider the case that $P(p) + P(n)$ is equal to 1, $E[(Y_i(1) - Y_i(0)) \cdot (M_i(1) - M_i(0))]$ is identifiable from the observed data when Assumptions 1 and 2 are met (i.e., $E[M_i(1) - M_i(0)] \neq 0$, $E[Y_i(1) - Y_i(0)] \neq 0$, M and X are unconfounded, Y and M are unconfounded if X is controlled). This is called the Path Average Causal Effect (PATE). That is the causal effects from X to Y through the path X→M→Y. PATE=ATE if Assumptions 4 and 5 are held.

Assumption 6 is vital for equation (6) to be held. In practice, we multiply the coefficient of the first-step regression ( $P(p) - P(n)$ ) and the second-step regression ( $E[(Y_i(1) - Y_i(0))]$ ). If heterogeneity do exist, $E[Y_i(1) - Y_i(0)] \neq E[Y_p(1) - Y_p(0)] \neq E[Y_n(1) - Y_n(0)]$. The calculated PATE is not equal to the theoretical one: $(P(p) - P(n)) \cdot E[Y_i(1) - Y_i(0)] \neq E[Y_p(1) - Y_p(0)] \cdot P(p) - E[Y_n(1) - Y_n(0)] \cdot P(n)$.

If Assumption 6 is violated, the bias can be calculated as

$$
\begin{aligned}
&E[Y_p(1) - Y_p(0)] \cdot P(p) - E[Y_n(1) - Y_n(0)] \cdot P(n) - (P(p) - P(n)) \\
&\qquad \cdot \{E[Y_p(1) - Y_p(0)] \cdot P(p) + E[Y_n(1) - Y_n(0)] \cdot P(n)\} \\
&= E[Y_p(1) - Y_p(0)] \cdot P(p)(1 - P(p) + P(n)) - E[Y_n(1) - Y_n(0)] \cdot P(n)(1 + P(p) - P(n)).
\end{aligned}
$$

Say

$$\omega = P(p) - P(n). \qquad (7)$$

Then, the bias is

$$\varepsilon = E[Y_p(1) - Y_p(0)] \cdot P(p) \cdot (1 - \omega) + E[Y_n(1) - Y_n(0)] \cdot P(n) \cdot (1 + \omega). \qquad (8)$$

We exclude the individuals whose $M_i(1) = M_i(0)$ because their treatment variable will not affect their outcome variable if all assumptions are met. Based on Assumption 4, the only way for X to affect Y is through M. Thus, if M takes the same values for all individuals, their average causal effect will be zero. In practice, it is easy to recognize these individuals and delete them from the overall analyses. If this is the case, the estimate will be local. If we do not exclude $M_i(1) = M_i(0)$, the estimates will be biased. The calculated ATE is shown below.

$$ATE_{cal} = (P(p) - P(n)) \cdot \{E[Y_P(1) - Y_P(0)] \cdot P(p) + E[Y_n(1) - Y_n(0)] \cdot P(n)\}.$$

If Assumption 6 holds, $E[Y_P(1) - Y_P(0)] = E[Y_n(1) - Y_n(0)]$. The calculated $E[Y_i(1) - Y_i(0)]_{cal} = E[Y_P(1) - Y_P(0)] \cdot P(p) + E[Y_n(1) - Y_n(0)] \cdot P(n)$, but $P(p) + P(n) = P(M_i(1) \neq M_i(0)) < 1$. Therefore, $E[Y_i(1) - Y_i(0)]_{cal} = (P(p) + P(n)) \cdot E[Y_P(1) - Y_P(0)] = (P(p) + P(n)) \cdot E[Y_n(1) - Y_n(0)]$.

Thus, a bias will be produced.

$$\begin{aligned}
&E\left[Y_p(1) - Y_p(0)\right] \cdot P(p) - E[Y_n(1) - Y_n(0)] \cdot P(n) - \left(P(p) - P(n)\right) \\
&\qquad \cdot \left\{E\left[Y_p(1) - Y_p(0)\right] \cdot P(p) + E[Y_n(1) - Y_n(0)] \cdot P(n)\right\} \\
&= E\left[Y_p(1) - Y_p(0)\right] \cdot P(p)\left(1 - P(p) + P(n)\right) - E[Y_n(1) - Y_n(0)] \cdot P(n)\left(1 + P(p) - P(n)\right) \\
&= E\left[Y_p(1) - Y_p(0)\right] \cdot P(p) \cdot (1 - \omega) + E[Y_n(1) - Y_n(0)] \cdot P(n) \cdot (1 + \omega) \\
¬e: E\left[Y_p(1) - Y_p(0)\right] = E[Y_n(1) - Y_n(0)]. \qquad (9)
\end{aligned}$$

This bias is the same as equation (8) in terms of form. However, $E\left[Y_p(1) - Y_p(0)\right] = E[Y_n(1) - Y_n(0)]$ in equation (9), but $E\left[Y_p(1) - Y_p(0)\right] \neq E[Y_n(1) - Y_n(0)]$ in equation (8). The bias of equation (8) is produced by $E\left[Y_p(1) - Y_p(0)\right] \neq E[Y_n(1) - Y_n(0)]$. The bias of equation (9) is produced by $P(p) + P(n) < 1$.

### *3.3 The Violation of Assumption 4*

For Assumption 4, we assume that all individual units follow the only path X→M→Y. However, there are also other paths that X can affect Y (also for all units), if Assumption 4 is violated. In such a case, the effect from other paths can be written as

$$O_i = Y_i(1|1 \; is \; for \; X_i) - Y_i(0|0 \; is \; for \; X_i). \; path: X \to Y.$$

What we should carefully notice is that $Y_i(1|1 \; is \; for \; X_i) - Y_i(0|0 \; is \; for \; X_i)$ is not equal to $Y_i(1) - Y_i(0)$ mentioned in 3.2. $Y_i(1) - Y_i(0)$ follows the function of the path M→Y. Thus, we can

write it as $Y_i(1|1\ is\ for\ M_i) - Y_i(0|0\ is\ for\ M_i)$. However, $Y_i(1|1\ is\ for\ X_i) - Y_i(0|0\ is\ for\ X_i)$ follows the function of another path: X→Y directly.

In such a case, we can also estimate the PATE accurately. However, PATE here is not equal to ATE because of the violation of Assumption 4.

In the FDC, we have two steps to obtain the final PATE. The first step is regressing X and M. This step is unaffected regardless of whether there are other paths from X to Y. The second step is regressing M and Y and controlling X at the same time. Controlling X will adjust all confounders between M and Y based on Assumption 2. Therefore, we can still obtain unbiased estimates for the PATE by multiplying the coefficients of X (the first-step regression) and M (the second-step regression). The only difference is that the coefficient of X in the second-step regression will be insignificant and approach 0 when there are no other paths. The coefficient will be significant when other paths do exist. However, we cannot simply conduct the weighted sum of the PATE and the effect of other paths. The reason is that the PATE is the unconfounded one, but the coefficient estimating for other paths (i.e., the coefficient of X in the second-step regression) is confounded.

The bias between the PATE and ATE is

$$E[O_i] = E[Y_i(1|1\ is\ for\ X_i) - Y_i(0|0\ is\ for\ X_i)].\ path: X \to Y. \tag{10}$$

However, this value is unrecognizable with the FDC. The FDC can only provide a biased estimate with confounding factors in the second step regression.

### 3.4 The Violation of Assumption 5

Assumption 5 is a worse version of Assumption 4. If there is only a local population following the path X→M→Y, we can only obtain the LATE. If the other population follows another path, the FDC estimates will be both local and path. If we can recognize the local population we can then estimate the LPATE for the local population following the specific path X→M→Y. If we fail to recognize them, then the estimates will be an unbiased ATE.

Say individual units i compromise the local population that follows the path X→M→Y. individual units j consists of the population whose X affects Y directly. P(i) is the proportion of the population i compared with the whole population. P(j) is the proportion of the other population. P(i)+ P(j)=1. Then,

$$C_i = Y_i(1|1\ is\ for\ M_i) - Y_i(0|0\ is\ for\ M_i). path: M \to Y.$$

$$N_j = Y_j\big(1|1\ is\ for\ X_j\big) - Y_j\big(0|0\ is\ for\ X_j\big).path{:}\ X \rightarrow Y.$$

As discussed in 3.2, if there are individuals whose $M_i(1) = M_i(0)$, our estimates will also be local even if all individuals follow the path X→M→Y (if we do not exclude them and calculate LATE, the ATE estimate will be biased). The $M_i(1) = M_i(0)$ can easily be recognized and excluded so that the unbiased LATE will be provided. For simplicity, we assume that all $M_i(1) \neq M_i(0)$, $M_j(1) \neq M_j(0)$. Therefore, in the following analyses, only the ATE is discussed. If there do exist $M_i(1) = M_i(0)$, $M_j(1) = M_j(0)$ for individual units i and j, we can simply exclude them and the following ATE will become LATE.

Case 1: Population j follows the same function as population i for X and M.

In such a case, the first-step regression is unaffected. The second-step regression can be divided into two parts. The first part is for the population i: $E[C_i]$. The second part is for the population j: $\frac{E[N_j]}{E[M_j(1)-M_j(0)]}$. The total causal effect of M→Y then becomes a weighted sum:

$$\underbrace{E[C_i] \cdot \mathrm{P}(i)}_{\text{Path: M} \rightarrow \text{Y for population i}} + \underbrace{\frac{E[N_j]}{E[M_j(1) - M_j(0)]} \cdot \mathrm{P}(j)}_{\text{Path: M'} \rightarrow \text{Y for population j, M' is virtual}} . \tag{11}$$

The second part $\frac{E[N_j]}{E[M_j(1)-M_j(0)]}$ is identifiable because the second step of regression will control X. Then, all confounders between M and Y are adjusted. To put it more vividly, this creates a new path (virtual: X→M'→Y) for the population j (M' is a virtual variable). The second regression captures the causal relationship of this new path (M'→Y), which is $\frac{E[N_j]}{E[M_j(1)-M_j(0)]}$. The confounder is adjusted by control X, making this part identifiable. Then, the only thing that needs to be done is weighted sum this part with the part of population i and multiplying the coefficients of the first-step regression.

$$ATE_{cal} = \underbrace{E[M_{i+j}(1) - M_{i+j}(0)]}_{\text{Path: X} \rightarrow \text{M (M') for population i (j)}} \cdot \underbrace{\{E[C_i] \cdot \mathrm{P}(i) + \frac{E[N_j]}{E[M_j(1) - M_j(0)]} \cdot \mathrm{P}(j)\}}_{\text{Path: M (M')} \rightarrow \text{Y for population i (j)}} .$$

$E[M_{i+j}(1) - M_{i+j}(0)] = E[M_i(1) - M_i(0)] = E[M_j(1) - M_j(0)]$ because population j follows the same function as population i for X and M. Then,

$$ATE_{cal} = \underbrace{E[M_i(1) - M_i(0)] \cdot E[C_i] \cdot \mathrm{P}(i)}_{\text{ATE for population i}} + \underbrace{E[N_j] \cdot \mathrm{P}(j)}_{\text{ATE for population j}} . \tag{12}$$

Thus, $ATE_{cal}$ is equal to the ATE. $E[M_i(1) - M_i(0)]$ is not written as 1 or -1 because monotonicity is not required. It is equal to $P(p) - P(n)$. $P(p)$ is the proportion of individuals of p compared to the population i and $P(n)$ is the proportion of individuals of n (see also 3.2).

Case 2: Population j follows the different functions from population i for X and M.

In such a case, the first-step regression is affected. It can be written as

$$E[M_{i+j}(1) - M_{i+j}(0)] = E[M_i(1) - M_i(0)] \cdot \mathrm{P}(i) + E[M_j(1) - M_j(0)] \cdot \mathrm{P}(j).$$

The second-step regression can be divided into two parts. The first part is for the population i: $E[C_i]$. The second part is for the population j: $\frac{E[N_j]}{E[M_j(1)-M_j(0)]}$. The total causal effect of M→Y then becomes a weighted sum: $E[C_i] \cdot \mathrm{P}(i) + \frac{E[N_j]}{E[M_j(1)-M_j(0)]} \cdot \mathrm{P}(j)$. It is the same as the equation for Case 1.

Then, the calculated ATE is

$$ATE_{cal} = \underbrace{\{E[M_i(1) - M_i(0)] \cdot \mathrm{P}(i) + E[M_j(1) - M_j(0)] \cdot \mathrm{P}(j)\}}_{\text{Path: X} \rightarrow \text{M (M') for population i (j)}} \cdot \underbrace{\{E[C_i] \cdot \mathrm{P}(i) + \frac{E[N_j]}{E[M_j(1) - M_j(0)]} \cdot \mathrm{P}(j)\}}_{\text{Path: M (M')} \rightarrow \text{Y for population i (j)}}. \quad (13)$$

The true ATE should be

$$ATE_{cal} = \underbrace{E[M_i(1) - M_i(0)] \cdot E[C_i] \cdot \mathrm{P}(i)}_{\text{ATE for population i}} + \underbrace{E[M_j(1) - M_j(0)] \cdot \frac{E[N_j]}{E[M_j(1) - M_j(0)]} \cdot \mathrm{P}(j)}_{\text{ATE for population j}}. \quad (14)$$

We can see that the main issues lie in the first step of regression. The i and j do not follow the same function between X and M, which leads to a large bias in the coefficient estimation. $E[C_i] \cdot \mathrm{P}(C_i)$ and $\frac{E[N_j]}{E[M_j(1)-M_j(0)]} \cdot \mathrm{P}(N_j)$ should have multiplied their own first-step coefficients, which are $E[M_i(1) - M_i(0)]$ and $E[M_j(1) - M_j(0)]$. However, they both multiply the actual first-step regression coefficient $E[M_{i+j}(1) - M_{i+j}(0)]$. In Case 1, $E[M_{i+j}(1) - M_{i+j}(0)] = E[M_i(1) - M_i(0)] = E[M_j(1) - M_j(0)]$ because population j follows the same function as population i for X and M. Thus, this issue can be neglected. However, this issue still holds in Case 2.

The bias can be calculated as

$$ATE - ATE_{cal} = E[C_i] \cdot \mathrm{P}(i) \cdot \mathrm{P}(j)\{E[M_i(1) - M_i(0)] - E[M_j(1) - M_j(0)]\} + \frac{E[N_j]}{E[M_j(1) - M_j(0)]} \cdot \mathrm{P}(i) \cdot \mathrm{P}(j) \cdot \{E[M_j(1) - M_j(0)] - E[M_i(1) - M_i(0)]\}.$$

Say

$$\gamma = \mathrm{P}(i) \cdot \mathrm{P}(j) \cdot \{E[M_i(1) - M_i(0)] - E[M_j(1) - M_j(0)]]. \quad (15)$$

Then, the bias is

$$\varepsilon = \gamma \cdot \{E[C_i] - \frac{E[N_j]}{E[M_j(1) - M_j(0)]}\}. \tag{16}$$

In such a case, the estimate of the FDC is neither ATE nor PATE, but a value with a bias of $\varepsilon$ compared with the ATE.

The above analyses are binary-based, but they are all linear derivations. Therefore, the results can be generalized to any continuous or discrete variables and can be linked to the linear model that is often used in economics and social science.

## 4. Comparisons of the front-door criterion with the instrumental variables

### *4.1 Comparisons of Assumptions*

Both methods need to follow the SUTVA.

Random Assignment versus Unconfoundedness (ideas for adjusting confounders).

In the IV, random assignment is a very important assumption. They require the IV must be randomly assigned or at least as good as produced by a random assignment. This can help the IV avoid the effect of confounders so that the true causal effect can be given. In the FDC, the confounder is controlled by the assumption that M is unconfounded with X and there is no confounder between M and Y if we control X. Random assignment and unconfoundedness are two prerequisites to guarantee that the causal effect can be estimated by the IV and FDC.

Nonzero Average Causal Effect of Z on D versus Identifiability (Validity of the chosen variables).

The IV (Z) is required to be strongly correlated to the treatment variable (D). The philosophy of IV is to isolate a clean part of the treatment variable that is only driven by IV, and then calculate the causal effect of this clean part with the outcome variable. If the relationship between IV and treatment variables is too weak, the clean part will be invalid. Identifiability is a similar thing. However, due to the special position of M (between X and Y), we require the correlation between X and M, Y and M both significant.

Exclusion Restriction versus Uniqueness and Universality (sensitivity if violating key assumptions).

If the exclusion restriction is violated, the IV will provide a biased estimation. In such a case, even LATE estimates are biased (see Angrist, Imbens, and Rubin (1996)). The original assumption of FDC put a very strict assumption to guarantee the ability to catch ATE. However, this greatly restricts its usage. Our analyses in 3.3 and 3.4 prove that even if we violate these two assumptions, the FDC can still provide useful insight into causal relationships. If assumption 4 is violated, we can still provide unbiased PATE estimates. When assumption 5 is violated, if the group following other paths follows the same function of X and M as that of the original group, then unbiased ATE will be provided. However, if they do not follow the same function, then biased estimates will be shown.

Monotonicity versus No Heterogeneity (sensitivity if violating key assumptions).

In the IV, the effect of IV on treatment variables must be in one direction for all individuals (no defier), or IV estimations will be biased (Angrist, Imbens, and Rubin (1996)). For the FDC, this rule does not need to be followed because of the nature of mediator variables. However, no heterogeneity must be followed, or the estimation will also be biased (see 3.2).

### *4.2 Comparisons of Causal Estimations*

The IV estimates are LATE. This holds the issue of external validity. Of course, some solutions have been developed using marginal treatment effect (MTE) (e.g., Kowalski (2016)). However, IV never resolved the issues of PATE. Although the IV simulates a randomized controlled trial (RCT), nature may have conducted many RCTs. That is the treatment variable is affected by many factors (e.g., two IVs). However, the IV can only obtain the causal effect of the treatment variable driven by a selected instrumental variable (PATE) rather than the overall causal effect (ATE). That is the impact of X on Y through a special path (IV→X (driven by IV)→Y). Therefore, most of the IV estimates are both local and path: LPATE. The estimates will be biased if there are other paths (e.g., IV1→X→Y, IV2→X→Y). However, this seems to happen often in the real world, and it seems unsolvable for the IV method.

This issue also applied to the FDC. Although Assumptions 4, 5, and 6 seem to solve everything and can estimate ATE (PATE=ATE if Assumptions 4, 5, and 6 are met). However, this is only an assumption (it is too strong). You can hardly find a wonderful mediator who matches these assumptions perfectly. However, this paper proves that the FDC will provide unbiased PATE and ATE if Assumption 4 and Assumption 5 (Case 1) are violated, respectively. Therefore, even if some assumptions are violated, it can still provide parameter estimates that are equal to or even better than the IV (at least theoretically). Of course, the biased estimations will be provided if other assumptions are violated (Assumption 5 Case 2; Assumption 6). This is similar to the biased estimations of the IV if exclusion restriction and monotonicity are violated.

## 5. Applying the front door criterion in linear systems

Linear models are the most commonly used method for calculating conditional expectations in economics. Therefore, this paper explores the performance of the FDC in linear systems and how biased estimation it will provide when the key assumptions are violated. In this section, I tested the FDC using simulated data generated by various linear systems.

I used the models from each DAG to generate data (Figure 1 and 2). For each model, the data for exogenous variables are randomly produced by AI Copilot for 200 individual units. Then, the other variables' values were generated based on the diagram and exogenous variables.

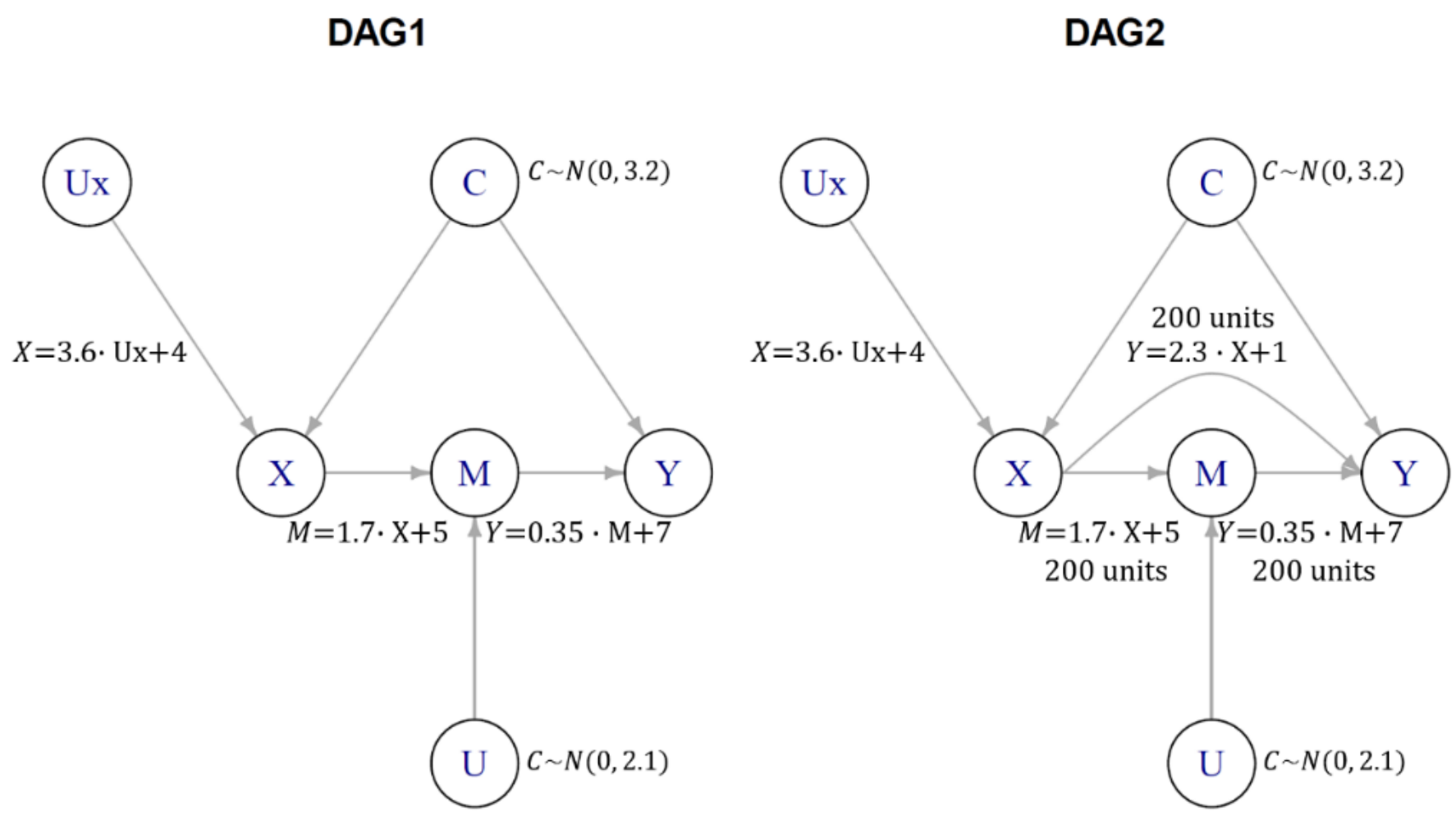

**Figure 1.** The DAGs 1 and 2 of the simulated data for testing.

The first model (DAG 1) follows all assumptions of the FDC (Fig. 1). For the first model, the FDC provides non-biased estimates (Table 1). Two-step regression both catch the causal effects (1.7065 and 0.3483). In the second regression, the coefficient of X is insignificant and approaches zero (0.0029). The second model violates uniqueness. All 200 units have two paths from X to Y. In such a case, the FDC also catches PATE (the same as the first model 1.7065·0.3483). As the theory part suggests, the coefficient of X is significant in this situation (2.3029). However, this cannot be regarded as the causal effect of the direct path (X→Y) because the confounder C is not adjusted for this path (the confounder for the path X→M→Y is adjusted by controlling X in the second regression). The estimate (2.3029) is very close to the true value of DAG 2 (2.3). The reason may be that the effect of the confounder is not that large in this model.

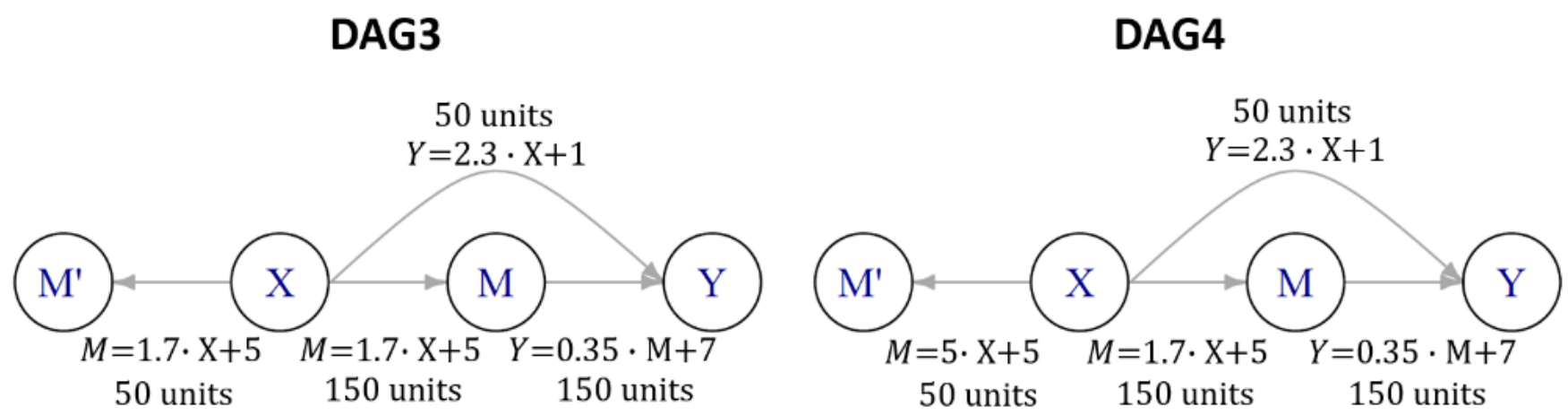

**Figure 2.** The DAGs 3 and 4 of the simulated data for testing.

I then tested the violation Assumption 5. DAG 3 is Case 1. That is the functions between X and M are the same for the group follows the path X→M→Y (150 units) and the path X→Y (50 units). The same as the theoretical analyses, the first-step regression is unaffected. The second-step regression equals to catching the effect of $E[C_i] \cdot \mathrm{P}(i) + \frac{E[N_j]}{E[M_j(1)-M_j(0)]} \cdot \mathrm{P}(j)\ (binary\ case)$ (see also 3.3).

Multiplying the two-step estimates 1.7000 and 0.6493 is approximately the model parameters $2.3 \cdot 0.25 + 1.7 \cdot 0.35 \cdot 0.75$. However, if it is Case 2 (DAG 4), the estimates will be biased. In this case, the functions between X and M are different for the groups following two paths. The first step coefficient is 2.6950 (far away from 1.7). This can be regarded as $E[M_i(1) - M_i(0)] \cdot \mathrm{P}(i) + E[M_j(1) - M_j(0)] \cdot \mathrm{P}(j)\ (binary\ case)$ in theoretical analyses (see 3.4). The second step coefficient 0.4623 is corresponding to $E[C_i] \cdot \mathrm{P}(i) + \frac{E[N_j]}{E[M_j(1)-M_j(0)]} \cdot \mathrm{P}(j)\ (binary\ case)$. Then, there will be a bias $\varepsilon$ between the multiplication of these two coefficients and the true ATE.

**Table 1.**The estimates for all four DAGs. signif. codes: 0 '***' 0.001 '**' 0.01 '*' 0.05 '.' 0.1 ' ' 1.

| Diagram | Step | Coefficients | Estimate | Std.Error | t value | Pr(>\|t\|) |
|---|---|---|---|---|---|---|
| DAG1 | Step 1 | Intercept | 4.5692 | 0.2511 | 18.2000 | <2e-16 *** |
| | | X | 1.7065 | 0.0029 | 589.6000 | <2e-16 *** |
| | Step 2 | Intercept | 7.1311 | 0.4919 | 14.4970 | < 2e-16 *** |
| | | M | 0.3483 | 0.0852 | 4.0890 | 6.3e-05 *** |
| | | X | 0.0029 | 0.1454 | 0.0200 | 0.984 |
| DAG2 | Step 1 | Intercept | 4.5692 | 0.2511 | 18.2000 | <2e-16 *** |
| | | X | 1.7065 | 0.0029 | 589.6000 | <2e-16 *** |
| | Step 2 | Intercept | 8.1311 | 0.4919 | 16.5300 | < 2e-16 *** |
| | | M | 0.3483 | 0.0852 | 4.0890 | 6.3e-05 *** |
| | | X | 2.3029 | 0.1454 | 15.8420 | < 2e-16 *** |
| DAG3 | Step 1 | Intercept | 5.0000 | 1.28E-14 | 3.91E+14 | <2e-16 *** |
| | | X | 1.7000 | 1.48E-16 | 1.15E+16 | <2e-16 *** |
| | Step 2 | Intercept | -1.4927 | 11.1676 | -0.1340 | 0.894 |
| | | M | 0.6493 | 0.0735 | 8.8320 | 5.51e-16 *** |
| DAG4 | Step 1 | Intercept | -5.6041 | 21.9931 | -0.2550 | 0.799 |
| | | M | 2.6950 | 0.2538 | 10.6200 | <2e-16 *** |
| | Step 2 | Intercept | -6.9002 | 0.7553 | -9.1360 | <2e-16 *** |
| | | M | 0.4623 | 0.0029 | 159.8670 | <2e-16 *** |

## 6. Conclusion

This paper redefines the key assumptions, explains the causal connotations, and deducts the estimated biases for the FDC, theoretically illustrating its power. Our analyses prove that the FDC can serve as a powerful tool for empirical research. Its original assumptions are super strict. However, we can still obtain useful insight into causal relationships even if some of them are violated. It holds a very subtle relationship with the IV. These two methods are comparable from all perspectives: assumptions, estimates, and biases if violating some key assumptions. Using the FDC and IV simultaneously may enable a more comprehensive insight into causal relationships.

## Acknowledgments

None.

## Declaration of interest statement

No potential conflict of interest was reported by the authors.

## Data Availability Statement

The data used is simulated based on DAGs from Fig. 1 and 2. I will make this available once the paper is published.